\documentclass[runningheads]{llncs}
\usepackage{chas}

\newcommand{\HEDAR}{Hilbert Edge Detection and Ranging}
\newcommand{\hedar}{HEDAR}
\newcommand{\Secref}[1]{Section~\ref{#1}} 

\begin{document}
\title{Combining Mathematical Morphology and the Hilbert Transform for Fully Automatic Nuclei Detection in Fluorescence Microscopy}%
\titlerunning{Mathematical Morphology and the Hilbert Transform for Nuclei Detection}

\author{Carl J. Nelson\inst{1}\orcidID{0000-0002-4114-1710} \and Philip T. G. Jackson\inst{2}\orcidID{0000-0001-7949-723X} \and Boguslaw Obara\inst{2}\thanks{\email{boguslaw.obara@durham.ac.uk}}}
\authorrunning{C. J. Nelson et al.}

\institute{University of Glasgow, Glasgow, UK \and Durham University, Durham, UK}
\maketitle
\begin{abstract}
    Accurate and reliable nuclei identification is an essential part of quantification in microscopy.
    A range of mathematical and machine learning approaches are used but all methods have limitations.
    Such limitations include sensitivity to user parameters or a need for pre-processing in classical approaches or the requirement for relatively large amounts of training data in deep learning approaches.
    Here we demonstrate a new approach for nuclei detection that combines mathematical morphology with the Hilbert transform to detect the centres, sizes and orientations of elliptical objects.
    We evaluate this approach on datasets from the Broad Bioimage Benchmark Collection and compare it to established algorithms and previously published results.
    We show this new approach to outperform established classical approaches and be comparable in performance to deep-learning approaches.
    We believe this approach to be a competitive algorithm for nuclei detection in microscopy.
\end{abstract}

\section{Introduction}\label{sec:introduction}

Quantitative biology relies heavily on the analysis of fluorescence microscopy images.
Key to a large number of experiments is the ability to detect and count the number of cells or nuclei in an image or region of interest.
Frequently, such as for high-throughput and large-scale imaging experiments, this is the first step in an automated analysis pipeline.
 As such, the scientific community requires novel, robust and automatic approaches to this essential step.

Manual cell counting and annotation, by visual inspection, is difficult, labour intensive, time consuming and subjective to the annotator involved.
As such, a large number of groups have proposed a variety of approaches for automatic nuclei detection in fluorescent microscopy images.
In recent years proposed methods have included:
\begin{itemize}
  \setlength\itemsep{0pt}
  \item variations upon thresholding-based segmentation, which are limited in performance due to intensity inhomogeneity and nuclei/cell clustering~\cite{GPSetal2006};
  \item H-minima transforms~\cite{JK2010};
  \item voting-based techniques~\cite{XLM2014}, which both show good results but are sensitive to parameters;
  \item gradient vector flow tracking and thresholding~\cite{KWTetal2015,LLTetal2007}, such PDE-based methods require strong stopping and reinitialisation criteria, set in advance, to achieve smooth curves for tracking;
  \item using Laplacian of Gaussian filters~\cite{XLBetal2016}, which is low computational complexity but struggles with variation in size, shape and rotation of objects within an image; and
  \item graph-cut optimisation approaches~\cite{NCKetal2016}, which requires the finding of initial seed points for each nucleus.
  \item convolutional neural networks, which generally require copious amounts of manually labelled training data, and struggle to separate overlapping objects \cite{jackson2017avoiding}
\end{itemize}

In our experience, nuclei size (scale) is an important contributor to
\begin{enumerate*}
  \item whether or not an algorithm is successful
  \item how robust an algorithm is to image variation and
  \item to the running time of an algorithm.
\end{enumerate*}
We have also found that many excellent algorithms exist for detecting small blobs, on the scale of a few pixels diameter, that fail or are less reliable for medium or large blobs, \ie{} over ten pixels diameter, when such objects tend to display significant eccentricity.
At this and larger scales the number of algorithms available decrease or make some assumptions on the data.

In this paper we apply a new ellipse detection algorithm based on mathematical morphology and the Hilbert transform to the task of counting medium blobs, \ie{} nuclei, on the scale of \numrange{10}{50} pixels in fluorescence microscopy images.
This new algorithm requires no preprocessing and no training.
Simplistically, the algorithm performs three steps:
\begin{enumerate*}
  \item Mathematical morphology is used to create an object sparse search space over different orientations and scales;
  \item The Hilbert transform is used to search this space and range object edges and for determining the ellipse/nucleus geometry at all pixels of the image;
  \item Prominent ellipses/nuclei are extracted and post-detection pruning is used to eradicate over detection.
\end{enumerate*}
We call this method \HEDAR{} (\hedar{}). All code is available on GitHub at \url{https://github.com/ChasNelson1990/Ellipse-Detection-by-Hilbert-Edge-Detection-and-Ranging}.

In the rest of this paper we introduce \hedar{} (\Secref{sec:methods}) and compare this method to several traditional and new algorithms for blob/nuclei counting in fluorescence microscopy (\Secref{sec:results}).
We have used the Broad Bioimage Benchmark Collection~\cite{LSC2012} throughout this paper to ensure comparison with previously published results, which we have reported where appropriate.



\section{Methods}\label{sec:methods}

Our new algorithm, \HEDAR{}, is a three step process.
First, we erode the original 2D, greyscale image \(I\) with a bank of line structuring elements \(SE_{d,\theta}\) of different integer pixel lengths \(d \in [1,d_{\max}]\) and rotations \(\theta \in [0,180)\).
We then stack the resulting eroded images into a four dimensional array \(M\), such that \(M(x,y,d,\theta) = (I \ominus SE_{d,\theta})(x,y)\).
We note that in 2D pixel-space we can define the limit on the number of rotations as \(N_{\theta} = 2d\), \ie{} the perimeter of the bounding box of the half-circle covered by all (symmetric) line structuring elements of length \(d\).
Thus, for a given $x,y,\theta$ where the point $(x,y)$ lies inside a bright ellipse, $M(x,y,:,\theta)$ yields a morphological intensity signal that drops off at the distance $s$ from $(x,y)$ to the boundary of the ellipse along the direction $\theta$.

Next, we calculate the 1D Hilbert transform over all morphological signals.
The transform will be maximal at a step in the morphological signal from high to low (and minimal at a step from low to high).
By identifying the maxima in each signal we can range the distance \(s\) from the current pixel to the nearest edge along that angle.
Mathematically, we can define this as,
\begin{equation}
    s(x,y,\theta) = \argmax_d \left( \hilbert \left( M(x,y,d,\theta) \right) \right).
\end{equation}

The major axis \(a\) of any ellipse, \ie{} nucleus, centred at \((x,y)\) in the image will be,
\begin{equation}
    a(x,y) = \max_{\theta}(s(x,y,\theta)).
\end{equation}
We can concurrently identify the orientation \(\phi\) of any nucleus centred at \((x,y)\) in the image as,
\begin{equation}
    \phi(x,y) = \argmax_{\theta} (s(x,y,\theta)).
\end{equation}
We then extract the distance to any edge at \ang{90} to the major axis \(a\) to determine the minor axis \(b\) of any nucleus centred at \((x,y)\) in the image.
Using the Hilbert transform makes this process robust to noise and blurring.

Like other mathematical morphological image processing approaches, such as neuriteness in vessel enhancement, so far the algorithm has assumed the presence of an ellipse at all points (\cf{} assuming a vessel in neuriteness), the next step is to identify likely ellipses/nuclei.
To accomplish this we element-wise multiply the minor and major axes at every pixel.
This produces a heat-map with high prominence peaks at the centres of any possible nuclei.

The prominence map is masked with a minimum threshold size \(t\) for Hilbert steps in morphological signals.
This acts similarly to local thresholding of the image: as each step must be at least as large as the threshold, we only consider pixels where the possible nucleus is brighter than its local background by at least \(t\). This suppresses false positives caused by background noise.
After smoothing the masked prominence map with a \(3\times3\) pixels median filter we then identify the regional maxima.
Each maximum represents a nucleus and we determine the major axis, minor axis and orientation from previous steps.

In post-processing, we cycle through all detected ellipses, removing the more overlapped ellipse of any pair with intersection-over-union greater than \SI{50}{\percent}.

We report results on image sets BBBC001v1~\cite{CJLetal2006}, BBBC002v1~\cite{CJLetal2006}, BBBC004v1~\cite{RLSetal2008} and BBC039v1~\cite{CRGetal2018} of the Broad Bioimage Benchmark Collection~\cite{LSC2012}.



\section{Results}\label{sec:results}

\subsection{Combined Mathematical Morphology and Hilbert Transform for Accurate Nuclei Counting}

\begin{figure}
    \centering
    \includegraphics[width=\linewidth]{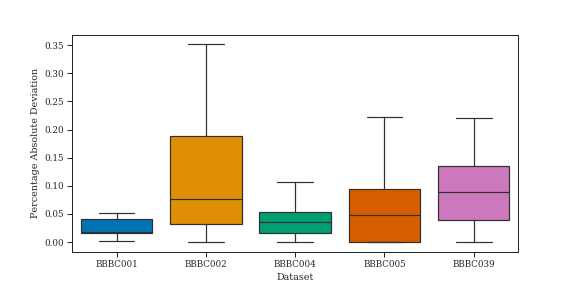}
    \caption{\textbf{\HEDAR{} enables accurate nuclei counting in fluorescent microscopy.} Distributions of percentage absolute deviation from known number of nuclei for \hedar{} across five fluorescence microscopy data sets with ground truth counts.}\label{fig:overview}
\end{figure}

Combined mathematical morphology and the Hilbert transform in our \HEDAR{} algorithm enables accurate nuclei counting in fluorescence microscopy, as show in \cref{fig:overview}.
Most nuclei are accurately counted giving a low mean percentage error.
\hedar{} does fail to detect a small number of cells, specifically irregular cells and some clustered or overlapping cells.

\begin{figure}
    \strut\hfill
    \begin{subfigure}{.5\linewidth}
    \includegraphics[width=\linewidth, clip, trim=200 175 100 125]{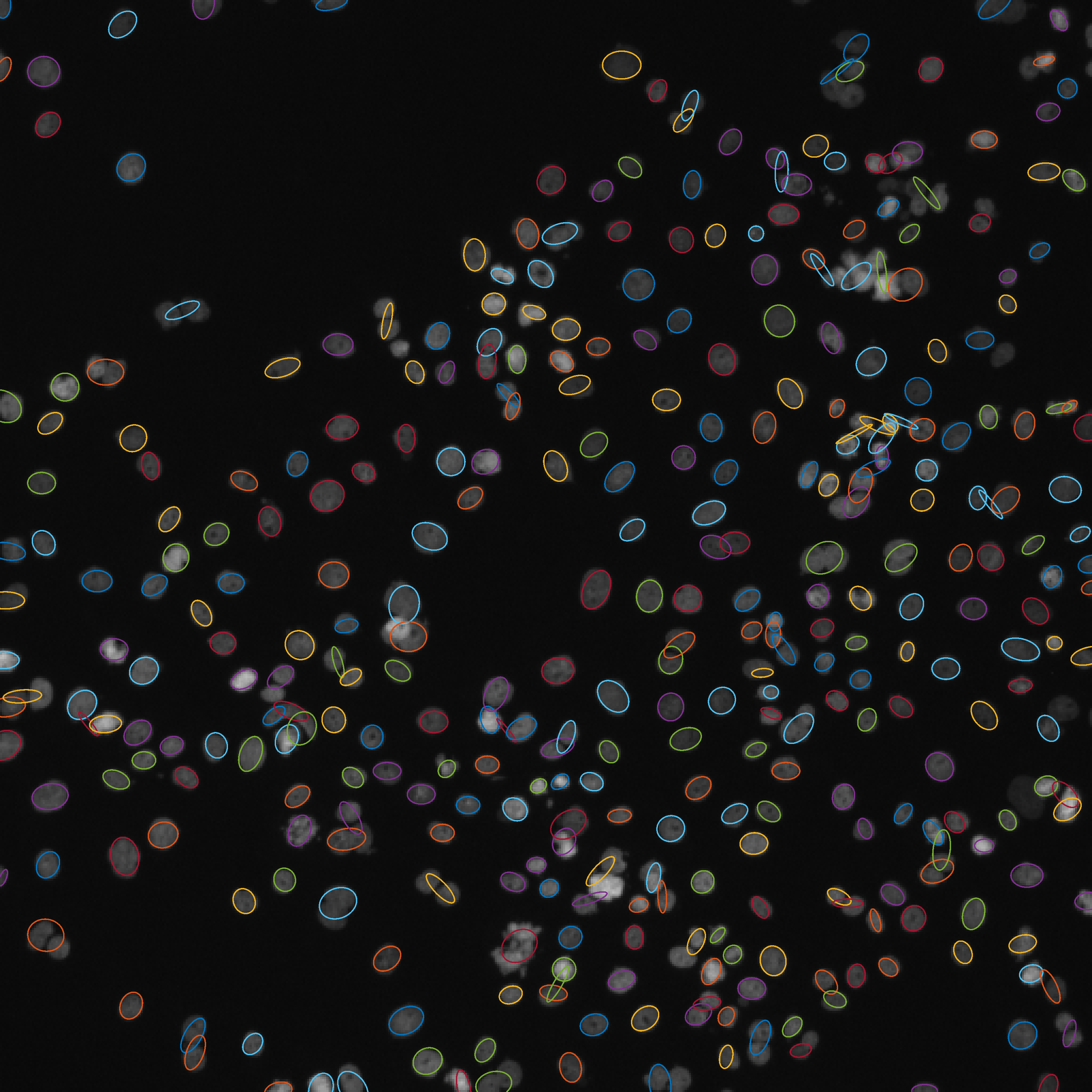}
    \caption{}\label{fig:humans:eg}
    \end{subfigure}
    \hfill
    \begin{minipage}{0.45\linewidth}
        \begin{subfigure}[b]{\linewidth}
            \centering
            \begin{tabular}{lr}
                \toprule
                Method & {Mean \% Error}\\
                \midrule
                Humans & 11\\
                \cite{CJLetal2006} & 6.2\\
                \midrule
                \hedar{} & 2.6\\
                \bottomrule
            \end{tabular}
            \caption{}\label{fig:humans:001table}
            \vspace{10pt}
        \end{subfigure}
        \begin{subfigure}{\linewidth}
            \centering
            \begin{tabular}{lr}
                \toprule
                Method & {Mean \% Error}\\
                \midrule
                Humans & 19\\
                \cite{CJLetal2006} & 17\\
                \midrule
                \hedar{} & 2.6\\
                \bottomrule
            \end{tabular}
            \caption{}\label{fig:humans:002table}
        \end{subfigure}
    \end{minipage}

    \begin{subfigure}{0.45\linewidth}
        \centering
        \includegraphics[width=\linewidth]{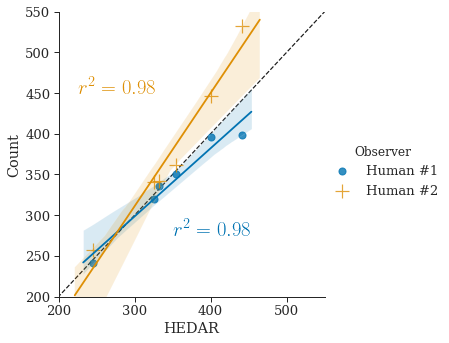}
        \caption{}\label{fig:humans:001plot}
    \end{subfigure}
    \begin{subfigure}{0.45\linewidth}
        \centering
        \includegraphics[width=\linewidth]{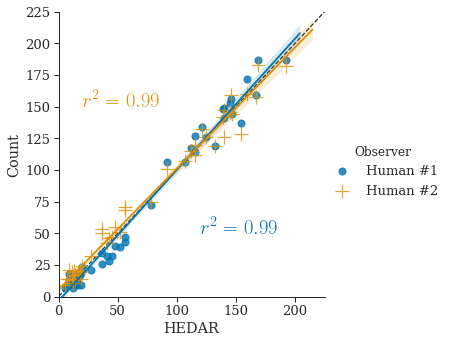}
        \caption{}\label{fig:humans:002plot}
    \end{subfigure}
    \hfill

    \hfill\strut
    \caption{\textbf{\HEDAR{} strongly agrees with human observers in counting nuclei in fluorescence microscopy images.} (\subref{fig:humans:eg}) Examples of successfully detected nulcei in the BBBC001 dataset of Hoechst 33342-labelled human HT28 colon cancer cells\cite{CJLetal2006}. (\subref{fig:humans:001table}) The percentage absolute deviation is reported for our \hedar{} algorithm and other published results on the BBBC001 dataset. The deviation between the two human observers is also reported. (\subref{fig:humans:002table}) The percentage absolute deviation is reported for our \hedar{} algorithm and other published results on the BBBC002 dataset. The deviation between the two human observers is also reported. (\subref{fig:humans:001plot}) Comparison of nuclei count from \hedar{} to the counts by two human observers on the BBBC001 dataset. (\subref{fig:humans:002plot}) Comparison of nuclei count from \hedar{} to the counts by two human observers on the BBBC002 dataset\cite{CJLetal2006}. }\label{fig:humans}
\end{figure}

By comparing the counting results of our automated algorithm to the counts by two human observers (see \Secref{datadetails} for details of datasets), we can show the accuracy of our \hedar{} algorithm (\cref{fig:humans:001plot} and \cref{fig:humans:002plot}).
The sample Pearson's correlation coefficient (\(r\) value) gives a measure of the correlation between the results of our automated \hedar{} algorithm and human observers --- an \(r\) value closer to \num{1} represents better correlation, \ie{} agreement.
For the BBBC001 datatset, the \(r\) value between the two human observers is \(r=0.92\).
With \(r\) values of \(0.98\) (with human observer \#1) and \(0.98\) (with human observer \#2) we clearly show that our \hedar{} algorithm has excellent performance for counting nuclei in fluorescent microscopy on dataset BBBC001.
Likewise, for the BBBC002 dataset we achieve \(r\) values of \(0.991\) and \(0.992\) (with human observer \#1 and \#2, respectively); the \(r\) value between the two human observers is \(0.992\).

This can be further compared using the mean percentage absolute deviation from the human mean, as reported by other papers (\cref{fig:humans:001table} and \cref{fig:humans:002table}).
We note that, as seen in~\cref{fig:overview}, the distribution of percentage absolute deviation is often skewed; as such, when comparing to full results from comparison methods we report the median and not the mean, which can be affected by outliers.

\begin{figure}
    \centering%
    \begin{subfigure}{0.5\linewidth}%
        \centering%
        \includegraphics[height=3cm]{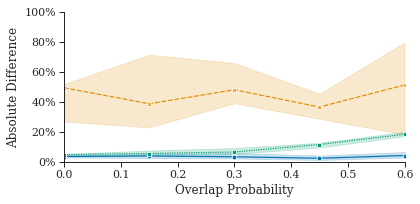}
        \caption{}\label{fig:results:bbbc004}%
    \end{subfigure}%
    \begin{subfigure}{0.5\linewidth}%
        \centering%
        \includegraphics[height=3cm]{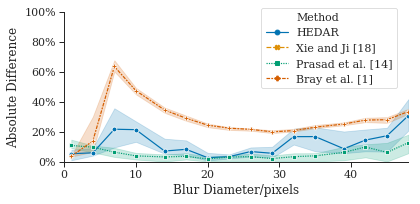}
        \caption{}\label{fig:results:bbbc005}%
    \end{subfigure}%
    \caption{\textbf{\HEDAR{} is able to cope with overlapping nuclei, increasing blur and out-of-focus images.} (\subref{fig:results:bbbc004}) Comparison of nuclei count from \hedar{} to the known counts for dataset BBBC004\cite{RLSetal2008} across the different overlap probabilities used to create the synthetic images. (\subref{fig:results:bbbc005}) Comparison of nuclei count from \hedar{} to the known counts for dataset BBBC005\cite{LSC2012} across the different blur diameters used to create the synthetic images.}\label{fig:results}%
\end{figure}

For completeness, we compared the count accuracy of our \hedar{} algorithm against all the Broad Bioimage Benchmark Collections datasets with 2D fluorescence microscopy images on fluorescently labelled nuclei (see \Secref{datadetails} for details of datasets).
These datasets also enable the exploration of how our \hedar{} algorithm is able to cope with overlapping nuclei (\cref{fig:results:bbbc004}), increasing blur (\cref{fig:results:bbbc005}) and relative focus (\cref{fig:results:bbbc005}).
Our \HEDAR{} algorithm consistently enables accurate counting of nuclei in fluorescence microscopy images under these challenging conditions.
\Cref{fig:results} shows the median percentage absolute difference compared to ground truth (BBBC004 and BBBC005; see~\Secref{datadetails}).

\subsection{\hedar{} is Able to Accurately Extract Ellipse Parameters from Detected Ellipses}

To investigate the accuracy of ellipse parameters using our \hedar{} algorithm we have emulated the synthetic experiments first shown in~\cite{FPC2014}.
We ran our method and two ellipse detection algorithms on synthetic images of a single ellipse without noise.
First, we compare to a well-cited, efficient, randomised elliptical Hough transform\cite{XJ2002}.
Second, we compare to a more recent but established non-iterative, geometric method for ellipse fitting: Ellifit\cite{PLC2012}.
Note that these algorithms are not optimised specifically for nuclei but have been shown to accurately extract nuclei in a variety of real world images; comparing to these methods in synthetic data sets a high standard for comparing our \hedar{} algorithm.

\begin{figure}
    \begin{subfigure}{\linewidth}%
        \centering
        \begin{tikzpicture}%
            \begin{groupplot}[
                group style={group size=3 by 1},
                width=\linewidth,
                xmin=0,
                xmax=1,
                x tick label style={rotate=90},
                xlabel=\(b/a\),
                ymin=3,
                ymax=29,
                y dir=reverse,
                scale only axis=true,
                width=2.5cm,
                enlargelimits=false,
                axis on top,
                ]%
                \nextgroupplot[
                    ylabel=\(a\),
                    title = \hedar{},
                    ]%
                    \addplot graphics[xmin=0,xmax=1,ymin=3,ymax=29] {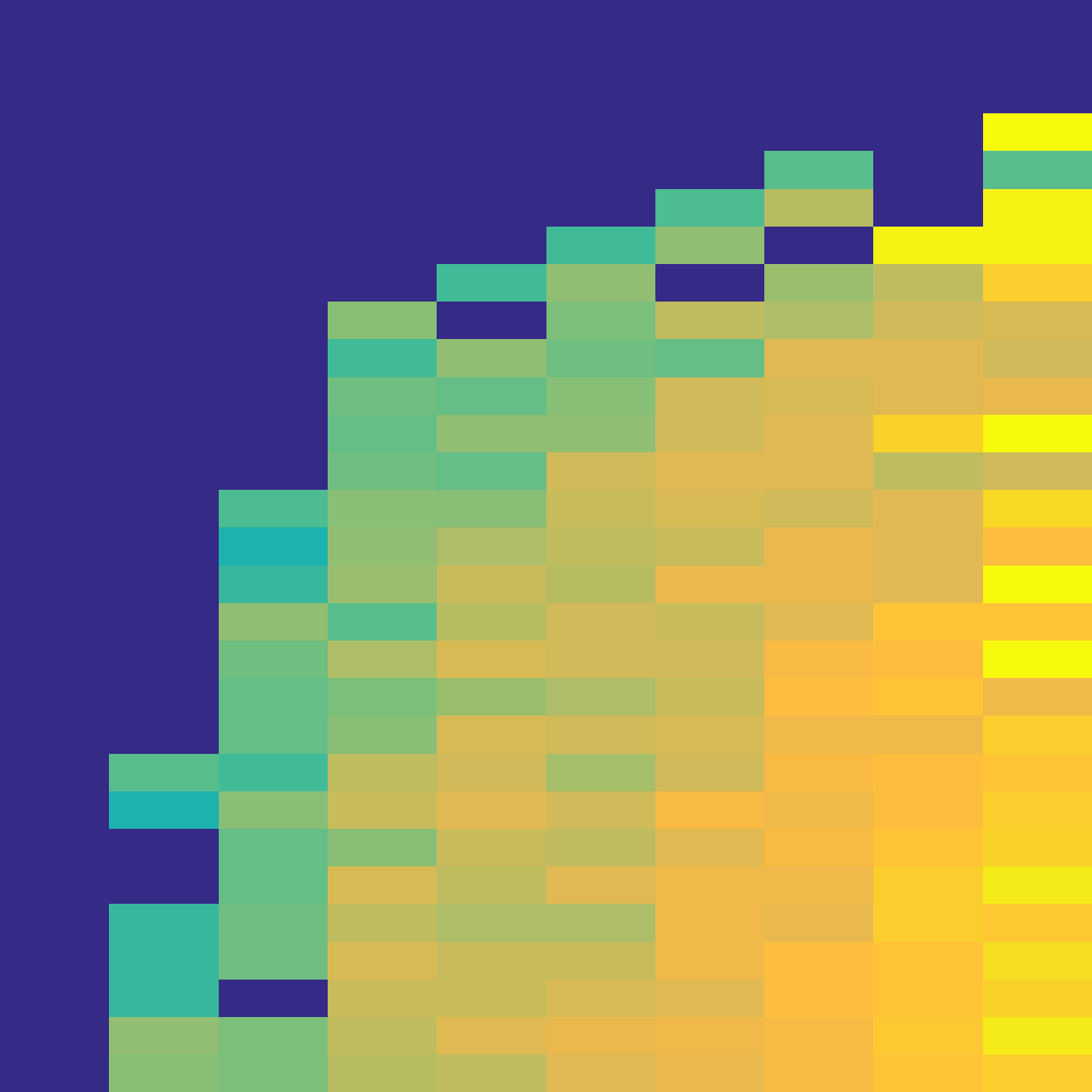};%
                \nextgroupplot[
                    ytick=\empty,
                    ylabel=\empty,
                    title = Xie and Ji\cite{XJ2002},
                    ]%
                    \addplot graphics[xmin=0,xmax=1,ymin=3,ymax=29] {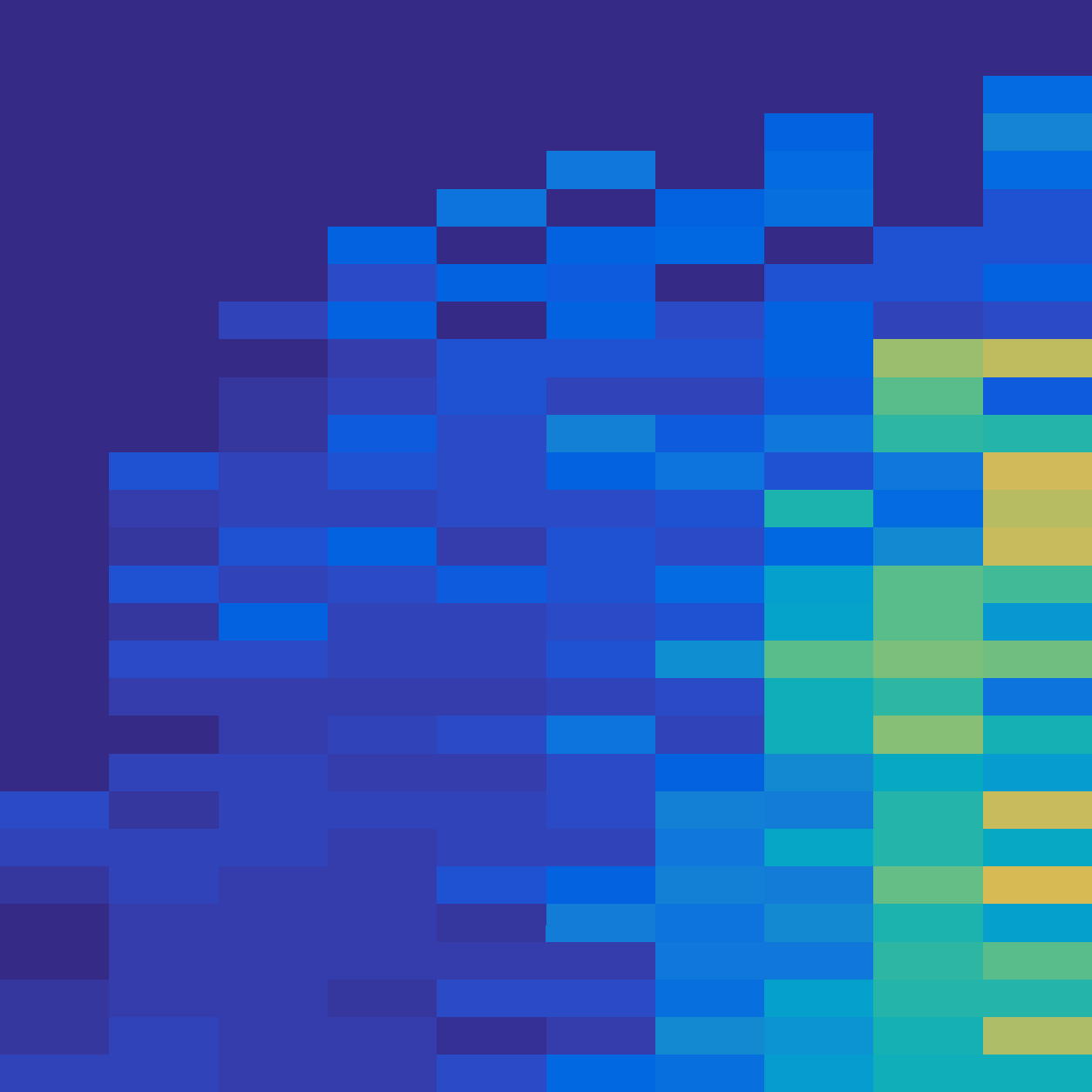};%
                \nextgroupplot[
                    ytick=\empty,
                    ylabel=\empty,
                    title = Prasad et al.\cite{PLQ2013},
                    ]%
                    \addplot graphics[xmin=0,xmax=1,ymin=3,ymax=29] {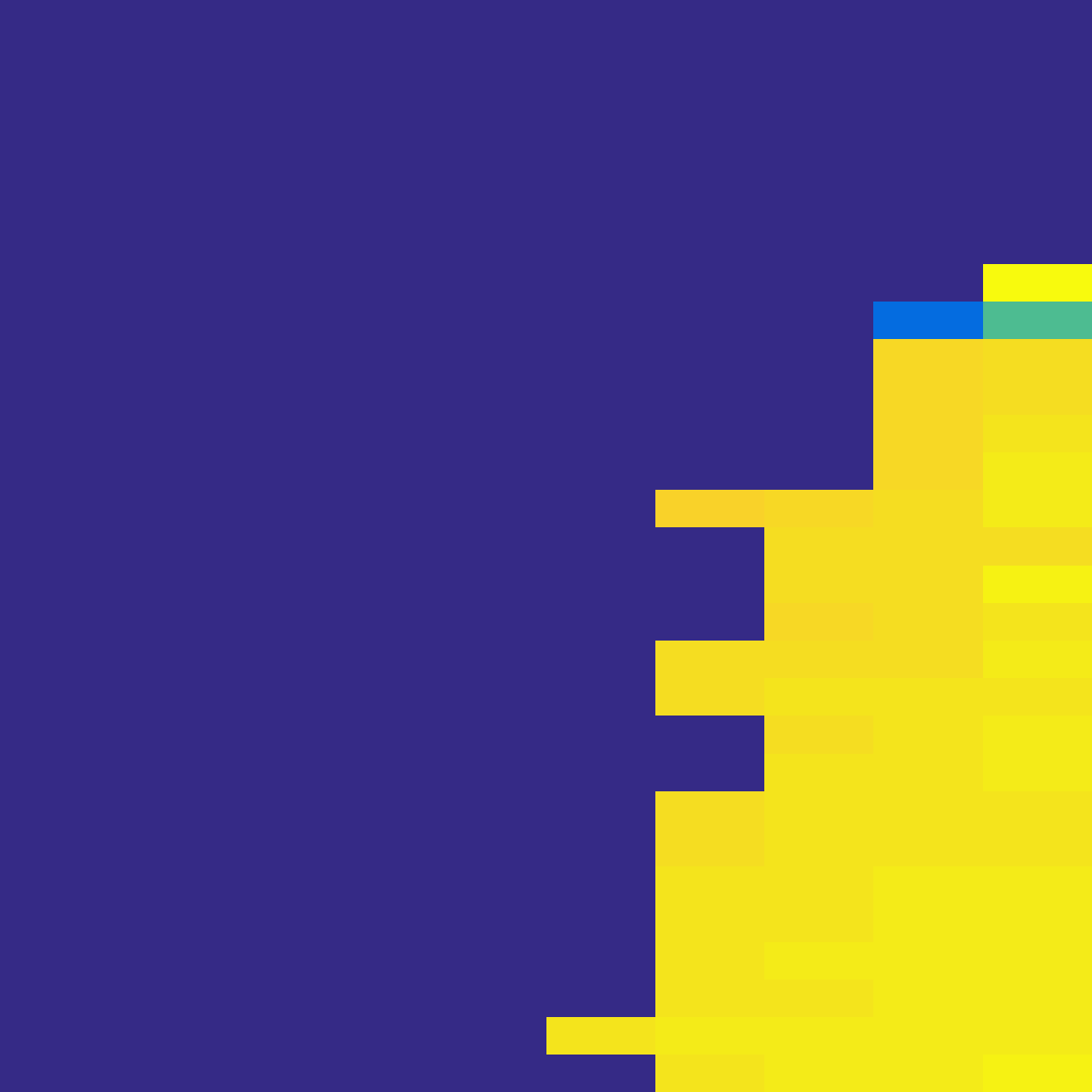};%
            \end{groupplot}
        \end{tikzpicture}%
        \caption{}\label{fig:accuracy:axes}%
    \end{subfigure}%

    \begin{subfigure}{\linewidth}%
        \centering
        \begin{tikzpicture}%
            \begin{groupplot}[
                group style={group size=3 by 1},
                width=\linewidth,
                xmin=0,
                xmax=1,
                x tick label style={rotate=90},
                xlabel=\(b/a\),
                ymin=0,
                ymax=179,
                y dir=reverse,
                scale only axis=true,
                width=2.5cm,
                enlargelimits=false,
                axis on top,
                ]%
                \nextgroupplot[
                    ylabel=\(\theta\),
                    title = \hedar{},
                    ]%
                    \addplot graphics[xmin=0,xmax=1,ymin=0,ymax=179] {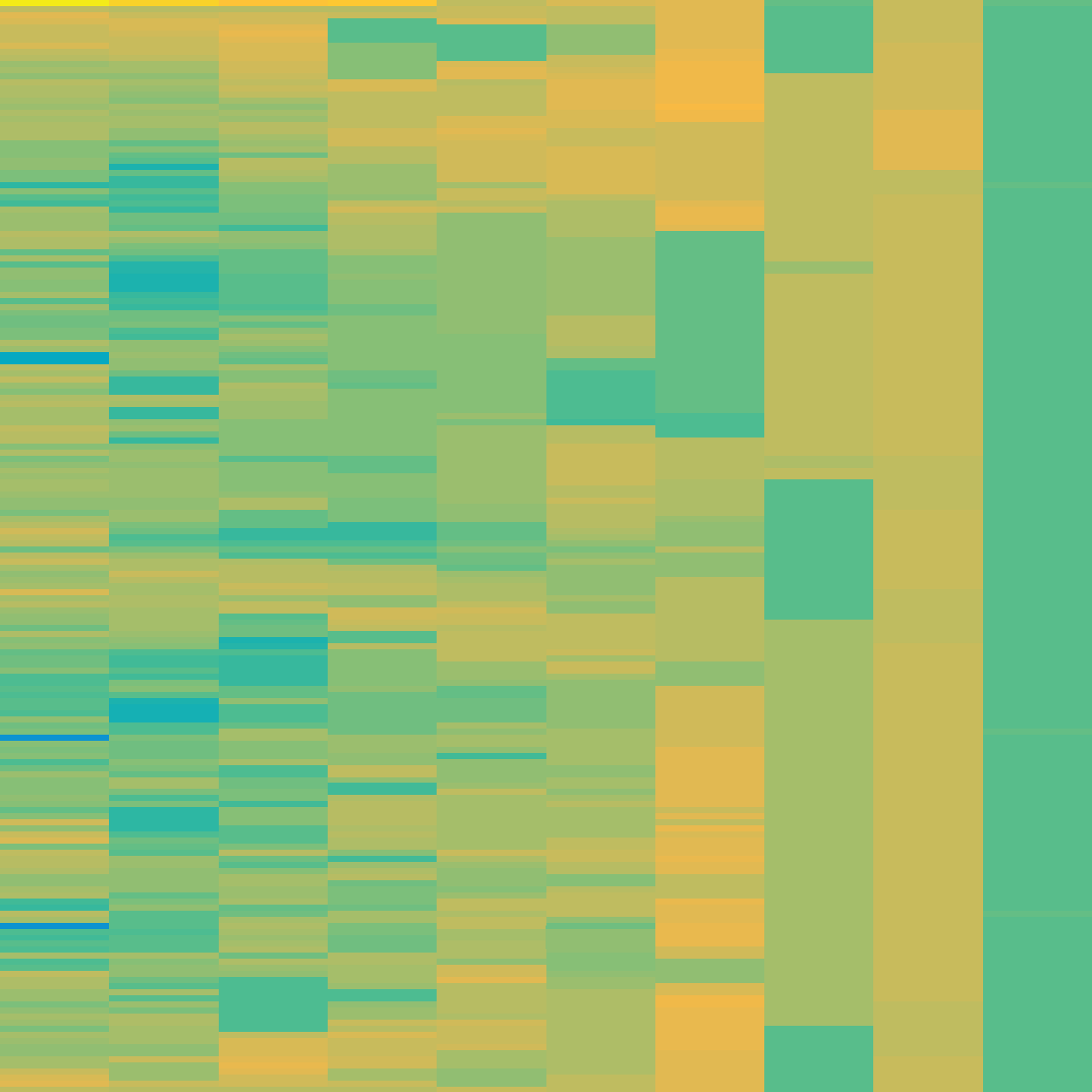};%
                \nextgroupplot[
                    ytick=\empty,
                    ylabel=\empty,
                    title = Xie and Ji\cite{XJ2002},
                    ]%
                    \addplot graphics[xmin=0,xmax=1,ymin=0,ymax=179] {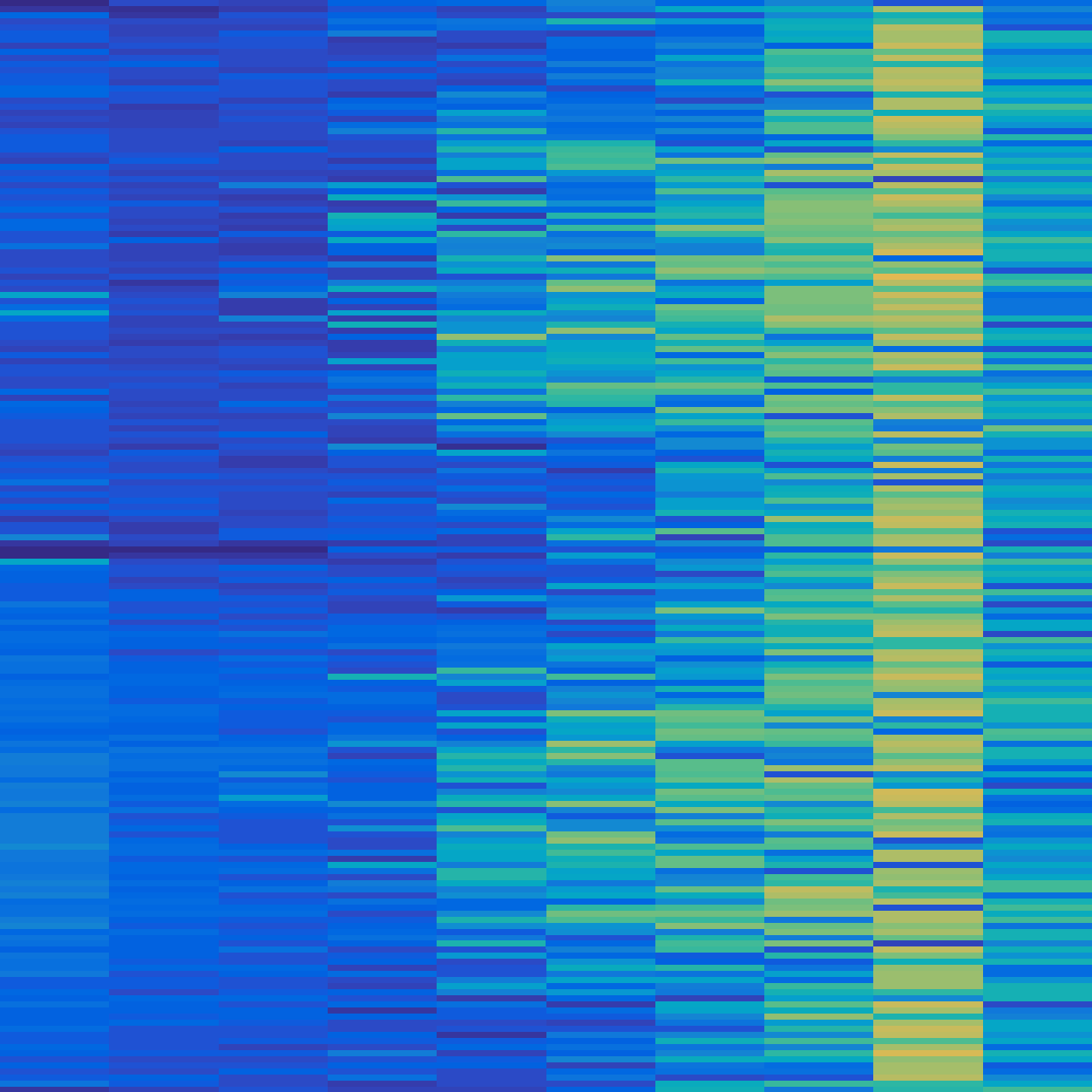};%
                \nextgroupplot[
                    ytick=\empty,
                    ylabel=\empty,
                    title = Prasad et al.\cite{PLQ2013},
                    ]%
                    \addplot graphics[xmin=0,xmax=1,ymin=0,ymax=179] {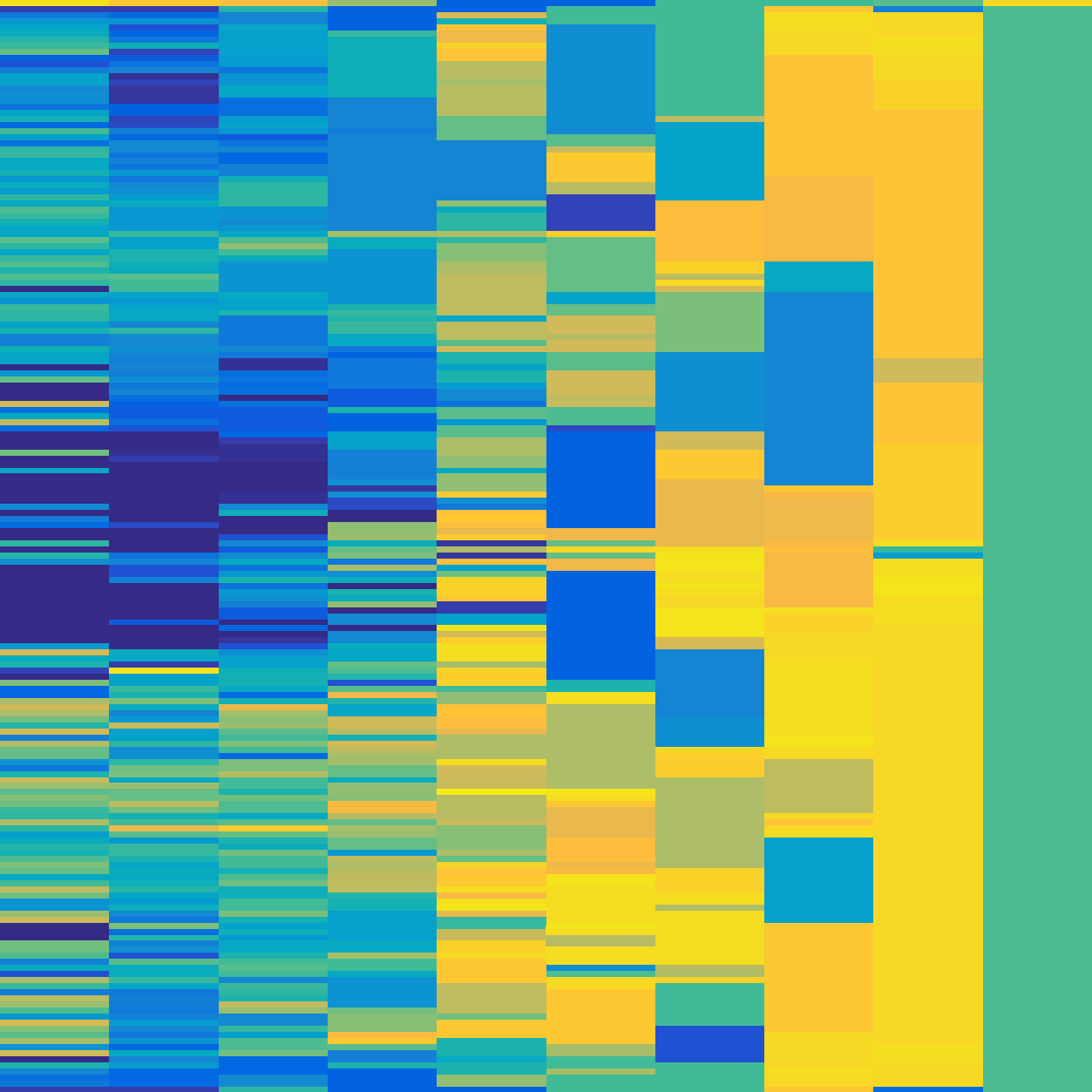};%
            \end{groupplot}
        \end{tikzpicture}%
        \caption{}\label{fig:accuracy:orientation}%
    \end{subfigure}%

    \vspace{5pt}
    {\centering
    \includegraphics[width=0.6\linewidth]{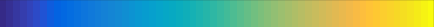}\\
    0.0 \hspace{0.245\linewidth} 0.5 \hspace{0.245\linewidth} 1.0\\
    Jaccard Similarity Index\\}
    \caption{\textbf{\HEDAR{} detects ellipses with high accuracy of axis lengths and orientation.} (\subref{fig:accuracy:axes}) Accuracy maps showing how \hedar{} and other ellipse detection algorithms cope with a range of major axis lengths \(a\) and axes ratios \(a/b\). (\subref{fig:accuracy:orientation}) Accuracy maps showing how \hedar{} and other ellipse detection algorithms cope with a range of orientations \(\theta\) and axes ratios \(a/b\).}\label{fig:accuracy}
\end{figure}

\Cref{fig:accuracy} shows how robust these methods are to size and eccentricity \cref{fig:accuracy:axes} as well as orientation (\cref{fig:accuracy:orientation}).
In these plots each `pixel' intensity value corresponds to the Jaccard similarity index~\cite{J1912} between the input image and the result of a given algorithm.
Hence, a yellow pixel (with a value closer to \num{1.0}) indicates that the given method was able to correctly identify the ellipse, \ie{} accurately extract ellipse parameters.
A deep blue pixel (with a value of \num{0.0}) indicates that no ellipse was detected or that more than one ellipse was detected, under which situations a Jaccard similarity of zero is assigned.
Any pixel between these two extremes can be considered to have found an ellipse with more (yellow) or less (blue) accurate parameters.
Note that the Jaccard similarity index was chosen as it satisfies the triangle inequality..

In~\cref{fig:accuracy:axes} each pixel represents the minimum Jaccard similarity for all ellipses of a given major axis length \(a\) and axis ratio (minor axis over major axis; \(b/a\)) over all rotations.
Although \hedar{} does get as consistently high a Jaccard similarity index as the Ellifit algorithm, \hedar{} covers a much greater range of major axis and axis ratios, more like the range covered by the elliptical Hough transform. These results shows that \hedar{} is able to accurately extract ellipse parameters over a wide range of ellipse sizes and axes ratios (\cf{} eccentricity).

In~\cref{fig:accuracy:orientation} each pixel represents the minimum Jaccard similarity for ellipses of a given orientation (\(\theta\)) and axis ratio \(b/a\) for \(a\leq32\) pixels.
\hedar{} shows consistently higher Jaccard similarity indices over small axes ratios (\ie high eccentricity) when compared to both the Ellifit and elliptical Hough algorithms.

\subsection{\hedar{} is Able to Accurately Count Ellipses}

\begin{figure}
	\centering
    \includegraphics[width=6cm]{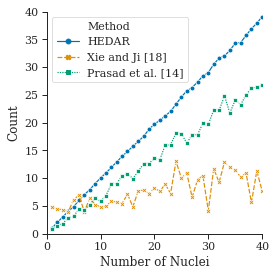}
	\caption{\textbf{\HEDAR{} can robustly detect the number of ellipses in an image.} Mean detected number of ellipses against actual number of non-overlapping, fixed-size ellipses for \hedar{} and comparator methods. Each point shows the mean of ten randomly generated synthetic scenarios.}\label{fig:proximity}
\end{figure}

Not only is the overall accuracy of any method important for its application but so too is the method's ability to deal with many objects.
\Cref{fig:proximity} shows how \hedar{} is able to correctly detect the number of ellipses in an image.
We ran \hedar{} on a set of synthetic images with a known number of small ellipses.
Each trial image was of of size \(256\times256\) and contained \(n\) small ellipses of varied size, position and orientation.
The plot shows how, as the number of ellipses increases, both the elliptical Hough transform and Ellifit struggle to correctly count the number of objects, whilst \hedar{} is able to maintain a correct count for most values of \(n\).

\subsection{Details of Benchmarking Data}\label{datadetails}

We have tested our new ellipse detection algorithm, \HEDAR{}, on five image datasets of fluorescent nuclei, all freely available from the Broad Bioimage Benchmark Collection~\cite{LSC2012}.
Full details of the datasets can be retrieved online but are summarised in~\cref{tab:data}.
All datasets are two dimensional.

Optimal parameters for \hedar{} were manually determined using a single image from each subset of data, \eg{} each z-focus in BBBC004, and those parameters were used to analyse each subset.
Parameters are also shown in~\cref{tab:data}.

\ctable[
  caption={BBBC datasets used to evaluate \hedar{} and compare to existing algorithms. Table indicates the number of images used from the dataset, whether the dataset was synthetic data (as opposed to real data) and whether the ground truth provide is a count of nuclei or a labelled, segmented image. Indicative \hedar{} parameters (\(d_{\max}\) and \(t\)) are also provided.},
  label=tab:data,
  width=\linewidth,
]{lccccclcc}{
  \tnote[1]{Although the whole dataset includes \num{9600} images of nuclei, only 20 randomly selected images per blur were used.}
  \tnote[2]{Nuclei count has been extracted from labelled ground truth.}
}{
    \toprule
    Accession & Citation          & Number          & Synthetic    & Truth     & Count                & Labels       & {\(d_{\max}\)} & {\(t\)} \\
    \midrule
    BBBC001v1 &\cite{CJLetal2006} & 6               &               & Human     & \checkmark            &               & 20    & 0.25  \\
    BBBC002v1 &\cite{CJLetal2006} & 50              &               & Human     & \checkmark            &               & 45    & 0.5   \\
    BBBC004v1 &\cite{RLSetal2008} & 100             & \checkmark    & Known     & \checkmark            & \checkmark    & 40    & 0.2   \\
    BBBC005v1 &\cite{LSC2012,BFHetal2012}     & 320\tmark[1]    & \checkmark    & Known     & \checkmark            & \checkmark    & 30    & 0.6   \\
    BBBC039v1 &\cite{CRGetal2018} & 50              &               & Human     & \checkmark\tmark[2]   & \checkmark    &       &       \\
    \bottomrule
}



\section{Conclusions}\label{sec:discussion}
These results show that \HEDAR{} is a competitive approach for nuclei counting and detection. \hedar{} performs at the same or better level of error than commonly used solutions and recent state-of-the-art solutions and gives results that agree with ground truth data with a high level of accuracy.

In this paper we have highlighted both the benefits of this method and those scenarios where this method may falter. We believe this information to be essential before users should use this method on their own data. We expect that further development of this method, such as optimisation for larger nuclei and adaptations to the Hilbert-edge ranging stages, will further improve the performance and the applicability of this technique.


\section{Acknowledgements}

During this work, CJN was supported by an EPSRC (UK) Doctoral Scholarship (EP/K502832/1).
PTGJ is supported by an EPSRC (UK) Doctoral Scholarship (EP/M507854/1).
The work in this paper was supported by an academic grant from The Royal Society (UK;~RF080232).
\bibliographystyle{splncs04}
\bibliography{ms}
\vfill\pagebreak
\end{document}